\documentclass[twocolumn,superscriptaddress,nofootinbib,aps,floatfix,preprintnumbers,amsmath,amssymb,groupedaddress]{revtex4}

\usepackage{amsmath}         
\usepackage{amssymb}         
\usepackage{amsfonts}        
\usepackage{graphicx}        
 
\usepackage{url}            
\usepackage{youngtab}	    
\usepackage{bbold}                

\graphicspath{{figures/}}	        

\renewcommand{\tilde}{\widetilde}   
\def\nn{\nonumber}

\let\oldenumerate\enumerate
\renewcommand{\enumerate}{
  \oldenumerate
  \setlength{\itemsep}{1pt}
  \setlength{\parskip}{0pt}
  \setlength{\parsep}{0pt}
}

\let\olditemize\itemize
\renewcommand{\itemize}{
  \olditemize
  \setlength{\itemsep}{1pt}
  \setlength{\parskip}{0pt}
  \setlength{\parsep}{0pt}
}

\newcommand{\antim}{{\fontsize{2}{3}\selectfont \Yvcentermath1 \yng(1,1)}}

\newcommand{\symsymrep}{{\fontsize{2}{3}\selectfont \Yvcentermath1 \yng(3)}}

\newcommand{\beq}{\begin{equation}}
\newcommand{\eeq}{\end{equation}}
\newcommand{\bea}{\begin{eqnarray}}
\newcommand{\eea}{\end{eqnarray}}

\begin{document}

\title{From S-confinement to 3D Chiral Theories: Dressing the Monopoles}

\author{Antonio Amariti}
\affiliation{Laboratoire de Physique Th\'eorique de l'\'Ecole Normale Sup\'erieure,
Paris 75005, France}

\author{Csaba Cs\'aki}
\affiliation{Department of Physics, LEPP, Cornell University, Ithaca, NY 14853, USA}

\author{Mario Martone}
\affiliation{Physics Department, University of Cincinnati, Cincinnati OH 45221-0011, USA}

\author{Nicolas Rey-Le Lorier}
\affiliation{Department of Physics, LEPP, Cornell University, Ithaca, NY 14853, USA}

\date{\today}

\begin{abstract}
Monopole operators play a central role in 3 dimensional supersymmetric dualities: a careful understanding of their spectrum is necessary to match chiral operators on either sides of a conjectured duality. In Chern-Simons theories ($k\neq0$), monopole operators acquire an electric charge, thus they need to be ``dressed'' by chiral matter superfields to be made gauge-invariant. Here we present strong evidence that ``dressed'' monopoles appear in $SU(N)$ chiral theories even for $k=0$ because of mixed CS terms generated along certain Coulomb branch directions. Our analysis is based on the dimensional reduction of 4-dimensional dualities which, for the simplest s-confining case, allows us to easily identify the spectrum of the electric chiral operators.
\end{abstract}

\maketitle

\section{Introduction}

Supersymmetry has had many important applications over the years: it is the leading candidate for solving the hierarchy problem, it allows a successful unification of gauge couplings, it predicts a potential dark matter candidate, and it is a necessary ingredient of string theories at high energy. However the arguably most successful application of supersymmetry has been 
 its use as a laboratory for testing non-perturbative physics effects and guessing dualities in various dimensions.

Supersymmetric dualities in 3D have been studied since the late 90s \cite{Aharony:1997bx,deBoer:1997kr,Aharony:1997gp,Karch:1997ux,Intriligator:1996mi,deBoer:1996mp,Hanany:1996ie,Kapustin:1999ha,Dorey:2000mirror}. One of their distinctive features is the role played by monopole operators which are non-trivially mapped across the dualities. Monopole operators are local disorder operators defined by requiring that the gauge field approaches a certain singular profile close to the point where the operator is inserted \cite{Borokhov:2002cg,Borokhov:2002ib}. These operators are commonly referred to as monopole operators because in Euclidean signature the gauge field singularity looks like that of a Dirac monopole or its non-Abelian generalization. 

In theories with Chern-Simons (CS) terms ($k\neq 0$), magnetically charged objects acquire an electric charge,  generically implying that a pure monopole operator $Y$ is no longer gauge invariant. Hence the right (chiral) operator that has to be matched across the duality is  ``dressed'' by some matter fields $\varphi$  such that $Y_{\text{dress}}\equiv Y_{\text{bare}}\varphi^{|k|}$, where $\varphi$ is an electrically charged operator.  Dressed monopoles are generic features of both $U(1)_k$ \cite{Intriligator:2013lca,Intriligator:2014fda} and $U(N)_k$ \cite{Aharony:2013dha,Aharony:2015pla} theories.

 Monopole operators have recently been systematically studied (e.g. \cite{Pufu:2013eda,Dyer:2013fja,Intriligator:2013lca,Aharony:2013dha,Intriligator:2014fda,Aharony:2015pla,Dyer:2015zha}). In particular in \cite{Aharony:2015pla} the super-conformal index was used to identify the chiral monopoles of generic $U(N)_k$ theories (with chiral and non-chiral matter) finding, surprisingly, that monopole operators are not always chiral. Furthermore the authors of \cite{Intriligator:2013lca,Intriligator:2014fda} have carefully investigated the relations between monopole operators and Coulomb branch (CB) operators. Many of their results will be used here.

The study of  monopole operators in $U(1)$ and $U(N)$ theories is made feasible by the extra {\it topological} global $U(1)_J$ symmetry  \cite{Aharony:1997bx}. Monopole operators are always charged under the $U(1)_J$, and are the only matter fields carrying such a charge, making them clearly identifiable. Simple gauge groups however don't have this extra global symmetry, making the study of the monopole operators much harder in $SU(N)$ theories.  This is the task that we will attempt to address here by investigating a concrete chiral 3D ${\cal N}=2$ $SU(N)$ model. 

Using dimensional reduction \cite{Aharony:2013dha} of 4D s-confining dualities \cite{Csaki:1996sm,Csaki:1996zb} we are able to obtain the low-energy description of 3D s-confining theories. These theories are described in the IR by very basic ``confined'' dynamics from which we can readily obtain the spectrum of the chiral operators of the UV theory and in particular study its monopole operators. By using this technique we can present evidence for the existence of ``dressed'' monopole operators in chiral $SU(N)$ theories at  zero CS level. A crucial ingredient of our novel findings is the presence of matter in the antisymmetric representation of $SU(N)$. These will generate mixed CS terms at one-loop along a particular unlifted $U(1)$ direction, which in turn induces the dressing of the monopole operator associated to that $U(1)$. 

This effect appears to be a generic feature of chiral theories with matter fields in tensor representations. Other theories with similar features will be discussed in a more comprehensive publication \cite{Amariti:2015sconf}.

\section{3D IR Duality in a Chiral $SU(6)$ Gauge Theory}

To make our discussion more concrete we perform the explicit analysis for a specific s-confining 3D theory.
At high energies the theory (referred to as the ``electric" theory) is a $(2+1)$ dimensional $\mathcal{N} = 2$ $SU(6)$ gauge theory with four fields in the antifundamental representation of the gauge group ($\bar{Q}$), two fields in the antisymmetric representation of the gauge group ($A$), and a vanishing superpotential. At low energies, this theory is described by a ``magnetic theory", with several gauge singlet ``meson-like"  fields and an s-confining superpotential. The matter and symmetry content of this theory is presented in Table \ref{tab:3dmatter}.

The low energy properties of this theory are derived from a similar $(3+1)$ dimensional s-confining theory through dimensional reduction. This procedure is by now fairly standard. It involves compactifying a dimension  on both sides of a 4D duality to obtain a duality between theories which live on $R^3 \times S^1$ and integrate out a flavor through a real mass deformation to decouple the extra super-potential term generated by a monopole configuration (KK monopole \cite{Lee:1997mo,Lee:1998mo,Lee:1998ca,Kraan:1998pi,Kraan:1998cal,Kraan:1998mon}) wrapping around the compactified dimension \cite{Aharony:2013dha}. The s-confining superpotential for the 3D theory can be obtained by removing from the superpotential of the 4D s-confining theory those fields which gain a mass when the flavor is decoupled \cite{Csaki:2014cwa}.

\begin{table}
\begin{align}
\begin{tabular}{l|c|ccccc}
&$SU(6)$&$SU(2)$&$SU(4)$&$U(1)_3$&$U(1)_4$&$U(1)_{R'}$\\
\hline
$\bar{Q}$&$\overline{\square}$&$1$&$\square$&-6&1&$0$\\
$A$&$\antim$&$\square$&1&3&0&$0$\\
 \hline
$b_1 \equiv  A \bar{Q}^2$&&$\square$&$\antim$&-9&2&$0$\\
$b_3 \equiv  A^3 $&&$\symsymrep$&1&9&0&$0$\\
$b_4 \equiv  A^4  \bar{Q}^2$&&1&$\antim$&0&2&$0$\\
$\tilde{M}_0\equiv Y$&&1&1&0&-4&$2$\\
$\tilde{M}_3\equiv A\tilde{Y}$&&$\square$&1&9&-4&2
\end{tabular}
\nonumber
\end{align}
\caption{Matter content of the 3D duality obtained from the 4D theory 
by applying the dimensional reduction procedure. \label{tab:3dmatter}}
\end{table}
 
Carrying out this procedure the 3D s-confining super-potential becomes (ignoring the overall scale): 
\begin{align}
W^{3D} = \tilde{M}_0\left(b_4^2+  b_3^2 b_1^2 \right) + \tilde{M}_3 \left( b_4 b_1 + b_3 b_1^2\right)~.
\label{WafterE}
\end{align}
The details of the real mass deformation can be found in Appendix \ref{RealMass}.

From Table \ref{tab:3dmatter} we can read off the chiral operators of the electric theory. The $b_i$'s are easily mapped to the gauge invariant meson-like operator of the electric theory  and it is tempting to associate $\tilde{M}_0$ and $\tilde{M}_3$ to unlifted CB directions parametrized by monopole operators. 
In the next section we will argue that while this is indeed the correct interpretation, the matching  of the $\tilde{M}_3$ field presents novel features due to the chiral nature of the theory.

\subsection{The Coulomb Branch Operators}

As was discussed in \cite{Csaki:2014cwa}, and will be discussed in more details in \cite{Amariti:2015sconf}, 
for an $SU(6)$ theory with a non-chiral field content of fundamental and antisymmetric fields the unlifted CB of the theory is described by three monopole operators, $Y$, $\tilde{Y}$, and $\hat{Y}$, associated with three unlifted $U(1)$ directions. These operators can be written explicitly in terms of the five ``fundamental'' monopole operators $Y_i$, $i = 1, ...,5$, as $Y \equiv Y_1 Y_2 Y_3 Y_4 Y_5$, $\tilde{Y} \equiv \sqrt{Y_1 Y_2^2 Y_3^2 Y_4^2 Y_5}$ and $\hat{Y} \equiv \left(Y_1 Y_2^2 Y_3^3 Y_4^2 Y_5 \right)^{1/3}$. In a chiral theory, however, we must be more careful. Alongside these directions, some matter fields acquire real masses and must be integrated out of the theory. The chiral nature of the theory allows the possibility that this could generate CS terms for the unbroken $U(1)$ gauge groups. These could in turn affect the equations of motions and change the structure of the moduli space. We examine all three CB operators in details. We will find that the $Y$ operator remains unlifted while $\hat{Y}$ is lifted. The operator $Y$ is readily associated to $\tilde{M}_0$. However the matching of the remaining $\tilde{Y}$ operator is less trivial. In fact, a CS term is generated along the $\tilde{Y}$ direction. This causes this operator to acquire an electric charge and to be ``dressed up'' into the correct gauge invariant operator, $A\tilde{Y}$. We will show with an explicit calculation that this is indeed the correct operator to describe the flat direction of the electric theory. Furthermore we will show that this operator has exactly the right abelian and non-abelian global quantum numbers to match the $\tilde{M}_3$ chiral operator leading to the association in Table \ref{tab:3dmatter}.

\subsubsection*{Generalities}

Here we present the essential ingredients to understand the CB dynamics of 3D $\mathcal N=2$ gauge theories. For reviews and more details see \cite{Aharony:1997bx,deBoer:1997kr,Aharony:2013dha,Csaki:2014cwa}.

A generic point of the CB of a $\mathcal N=2$ 3D $SU(N)$ gauge theory is parametrized by the VEV of the scalar component of the vector multiplet which, using gauge transformations, can be diagonalized. Thus a generic point of the CB can be parametrized by $N$ parameters: $\sigma_i$'s, $i=1,...,N$, satisfying $\sigma_1\geq\sigma_2\geq...\geq\sigma_N$ and $\sum_i\sigma_i=0$:
\begin{equation}\label{Mario1}
\varphi=
\left(
\begin{array}{cccc}
\sigma_1&0&...&0\\
0&\sigma_2&...&0\\
0&0&\ddots&0\\
0&0&...&\sigma_N
\end{array}
\right),\qquad\sum_{i=1}^N\sigma_i=0
\end{equation}
For generic values of the  $\sigma$'s $SU(N)\to U(1)^{N-1}$. Each unlifted $U(1)$ direction is parametrized by an operator $Y_i$, $i=1,..,N-1$, and thus semi-classically the theory has a $N-1$ dimensional CB. Each $Y_i$ is associated to one of the $N-1$ fundamental monopoles. 

At the quantum level, 3D instantons generate contributions to the super-potential \cite{Affleck:1982as} which, in absence of matter, fully lift the CB. In the presence of matter, however, the analysis becomes considerably more elaborate as the CB splits in different regions in which some of the quantum corrections are not generated because of the presence of fermionic zero-modes \cite{Aharony:1997bx,deBoer:1997kr,Aharony:2013dha,Csaki:2014cwa}. 

After accounting for non-perturbative corrections, we are left at most with operators describing unlifted $U(1)$ directions. For a Coulomb vacuum associated to a particular $U(1)_a$ to be allowed, it needs to satisfy\footnote{Here we assuming that both the tree level CS and FI terms vanish, $k=\xi=0$.}:
\begin{align}\label{cond}
&\sum_i 2\pi n^{(a)}_i\big|\Phi_i\big|^2=\xi^{(a)}_{eff}+\sum_b k^{(a,b)}_{eff}\sigma^{(b)},\\\label{cond1}
&\left(m_{\mathcal{R},i}+n^{(a)}_i\sigma^{(a)}\right)\Phi^{(a)}_i=0~,
\end{align}
where $n_i^{(a)}$ and $\sigma^{(a)}$ are, respectively, the charge of the $i$th field under and the scalar VEV for the unbroken $U(1)_a$, while $\Phi_i$ is the VEV of the $i$th matter field. The $k_{eff}^{(a,b)}$ are the CS terms generated at one-loop \cite{Intriligator:2013lca,Intriligator:2014fda}:
\begin{align}
k^{\left(a,b\right)}_{eff}&=\frac{1}{2}\sum_i n^{(a)}_i n^{(b)}_i\textrm{sign}(m_{\mathcal{R},i})~,
\label{AbelianCSTerm}
\end{align}
where $m_{\mathcal{R},i}$ is the real mass of the field $i$. Note that in all the cases we are interested in here, $m_{\mathcal{R},i}$ will be induced by a given $U(1)_{\bar{a}}$ VEVs, thus $m_{\mathcal{R},i}=n^{(\bar{a})}_i \sigma^{(\bar{a})}$. 
These formulae allows for both ``pure", $a=b$, and ``mixed", $a\neq b$, CS terms. A similar formula applies for the effective Fayet-Iliopoulos (FI) term
\begin{align}
\xi_{eff}^{(a)}&=\frac{1}{2}\sum_i n^{(a)}_i m_{\mathcal{R},i} \textrm{sign}(m_{\mathcal{R},i})~.
\label{FITerm}
\end{align}
Furthermore, CS terms will induce electric charges for the fields which are charged under the topological $U(1)_J$'s associated with the various $U(1)$ gauge factors:
\begin{align}
q_i^{(a)} = - \sum_b k^{(a,b)}_{eff}q_{J, i}^{(b)}~,
\label{InducedElectricCharges}
\end{align}
where $q_i^{(a)}$ is the electric charge under $U(1)_a$ generated for the $i$th field.

One loop CS terms can also be generated for non-abelian groups. Eq. \eqref{AbelianCSTerm} then  generalizes to:
\begin{align}
k^G_{eff}&=\frac{1}{2}\sum_{\textbf{r}_i}T_2(\textbf{r}_i)\textrm{sign}(m_{\mathcal R,\textbf{r}_i})~.
\label{NonAbCSTerm}
\end{align}
where the real masses $m_{\mathcal R,\textbf{r}_i}$ will again be induced by $U(1)$ VEVs.

\subsubsection*{The $Y$ Direction}
Far along on the moduli space, the $Y$ operator describes the direction diag$\left(\sigma,0,0,0,0,-\sigma \right)$. 
This field configuration spontaneously breaks $SU(6)$ to $SU(4) \times U(1)_1 \times U(1)_2$ where, up to an irrelevant normalization, the $U(1)_1$ and $U(1)_2$ factors are respectively associated with the generators diag$\left(1,0,0,0,0,-1\right)$ and diag$\left(2,-1,-1,-1,-1,2\right)$.
Under the unbroken gauge group the matter content of the theory decomposes as follows :
\begin{align}
\overline{\square} &\rightarrow \overline{\square}_{\left(0,1\right)} + \mathbb{1}_{\left(-1,-2\right)} +  \mathbb{1}_{\left(1,-2\right)}~, \nn \\
\antim &\rightarrow \antim_{\left(0,-2\right)} + \square_{\left(1,1\right)} +\square_{\left(-1,1\right)} + \mathbb{1}_{\left(0,4\right)}~.
\label{MatterDecomp}
\end{align}
We can now compute the possible CS terms generated by integrating out the heavy matter fields. From Eq.~(\ref{MatterDecomp}) we see that that all fields charged under $n^{(1)}$ come in pairs of fields whose contributions to the pure CS terms are equal and opposite: hence, all pure CS terms vanish. This argument does not apply to the mixed CS term however, which must be computed:
\begin{align}
k^{\left(1,2\right)}_{eff}=\frac{1}{2} \left(-8 -8 +8 +8 \right) =0~.
\label{MixedCSTerm}
\end{align}
So the mixed CS term also vanishes, but this is not a generic result and only occurs because of a non-trivial cancellation between the different contributions from the matter fields in the theory. All FI terms also vanish. It follows that this CB direction remains unlifted and unmodified by the chiral effects.
\subsubsection*{The $\tilde{Y}$ Direction and the Dressed Monopole}
Far along on the moduli space, the $\tilde{Y}$ direction describes the direction diag$\left(\sigma,\sigma,0,0,-\sigma,-\sigma \right)$.
This field configuration spontaneously breaks $SU(6)$ to $SU(2)_t \times SU(2)_m \times SU(2)_b \times U(1)_1 \times U(1)_2$. The subscripts ``$t$", ``$m$" and ``$b$" stand for ``top", ``middle" and ``bottom", and refer to the embedding of the $SU(2)$ subgroup of $SU(6)$ in the matrix representation of the 
group. Up to an irrelevant normalization, the $U(1)_1$ and $U(1)_2$ factors are respectively associated with the generators diag$\left(1,1,0,0,-1,-1\right)$ and diag$\left(1,1,-2,-2,1,1\right)$.
Under the unbroken gauge group the matter content of the theory decomposes as follows:
\begin{align}
\overline{\square} &\rightarrow \overline{\square}^t_{\left(-1,-1\right)} + \overline{\square}^m_{\left(0,2\right)} + \overline{\square}^b_{\left(1,-1\right)} , \nn \\
\antim &\rightarrow \mathbb{1}_{\left(2,2\right)} + \mathbb{1}_{\left(0,-4\right)} + \mathbb{1}_{\left(-2,2\right)} \nn \\
&+ \left(\square^t, \square^m\right)_{\left(1,-1\right)} + \left(\square^t, \square^b\right)_{\left(0,2\right)} + \left(\square^m, \square^b\right)_{\left(-1,-1\right)}~.
\label{MatterDecompYtilde}
\end{align}
 Once again, all pure CS terms will necessarily cancel because fields with non-vanishing $n^{(1)}$ terms necessarily come in pairs with opposite-sign real masses. This time, however, there is a non-vanishing mixed CS term:
\begin{align}
k^{\left(1,2\right)}_{eff}=\frac{1}{2} \left(-8 -8 +8 +8 -8 -8 \right) =-8~.
\label{MixedCSTermYtilde}
\end{align}
From Eq. (\ref{FITerm}), we see that there should also be an effective FI term generated for the $U(1)_2$ group, $\xi_{eff}^{(2)} = -8 \sigma$.
Hence \eqref{cond} and \eqref{cond1} are modified to:
\begin{align}
\sum_i 2\pi n_i^{(1)} |\Phi_i|^2 = k^{\left(1,2\right)}_{eff} \sigma^{(2)} = -8 \sigma^{(2)} ~,
\label{DFlatCond1}
\end{align}
\begin{align}
\sum_i 2\pi n_i^{(2)} |\Phi_i|^2 = k^{\left(1,2\right)}_{eff} \sigma^{(1)} + \xi_{eff}^{(2)}= -16 \sigma^{(1)}~,
\label{DFlatCond2}
\end{align}
where $\Phi_i$ represents the $i$-th massless field of the theory and the $\sigma^{(1),(2)}$ are the VEVs of the scalar component of the vector supermultiplets associated with the respective abelian gauge groups. In our notation, the ``undressed" $\tilde{Y}$ direction corresponds to $\sigma^{(1)} = \sigma >0$ while the matter fields' VEVs $\Phi_i$ and $\sigma^{(2)}$ are set to zero. This vacuum configuration is perfectly compatible with Eq. (\ref{DFlatCond1}) but it is inconsistent with Eq. (\ref{DFlatCond2}) due to the non-vanishing $k^{\left(1,2\right)}_{eff}$ and $\xi_{eff}^{(2)}$.
This condition can now only be satisfied if some VEVs for the matter fields are turned on.
For consistency with  Eq. \eqref{DFlatCond1} those fields must have $n_i^{(1)}=0$.
 Furthermore,  since the right-hand-side of Eq. (\ref{DFlatCond2}) is negative, those fields must also have $n_i^{(2)}<0$. From Eq. (\ref{MatterDecompYtilde}), we see that there is only one possible candidate: the $\mathbb{1}_{\left(0,-4\right)}$ field which was initially part of the field $A$ in the antisymmetric $SU(6)$ representation. Therefore, the effects of the chiral dynamics boil down to turning the CB parameter into a dressed monopole $A \tilde{Y}$. 
 As an independent check, we can verify that $A \tilde{Y}$ is in fact a gauge invariant operator. As explained above $A\to\mathbb{1}_{\left(0,-4\right)}$, thus it has charges $(0,-4)$ under the unbroken $U(1)$'s. From \eqref{InducedElectricCharges}, computing the $U(1)_2$ electric charge generated by the mixed CS term and the $U(1)_1$ magnetic charge, $\tilde{Y}$ has charges $(0,4)$ making the dressed operator gauge invariant. It is important to stress that gauge invariance of the dressed monopole implies the presence of the square root in the definition of the $\tilde{Y}$.

As we will show below, this dressed monopole also possesses  the correct global quantum numbers to match the field $\tilde{M}_3$ on the magnetic side of the duality. Note that this also includes the non-abelian flavor $SU(2)$ charge which $A_{\mathbb{1}_{(0,-4)}}\tilde{Y}$ inherits from the $A$ field.
All the one-loop non-abelian CS terms vanish:
\begin{align}
k^{SU(2)_t}_{eff}=\frac{1}{2} \left(4*(-1) + 2*2*(+1) \right) =0~, \nn \\
k^{SU(2)_m}_{eff}=\frac{1}{2} \left(2*2*(+1) + 2*2*(-1) \right) =0~, \nn \\
k^{SU(2)_b}_{eff}=\frac{1}{2} \left(4*(+1) + 2*2*(-1) \right) =0~.
\label{NonAbCSTermsYtilde}
\end{align}
The vanishing of all the non-abelian CS terms depends, once again, on a non-trivial cancellation which is realized for this specific matter content. In a generic chiral theory, these terms could be present.

\subsubsection*{The $\hat{Y}$ Direction}
Far along on the moduli space,  $\hat{Y}$ describes the direction diag$\left(\sigma,\sigma,\sigma,-\sigma,-\sigma,-\sigma \right)$.
This field configuration spontaneously breaks $SU(6)$ to $SU(3)_t \times SU(3)_b \times U(1)$. The subscripts ``$t$" and ``$b$" stand for ``top" and ``bottom", and refer 
 again to the position of the embedding of $SU(3)$ in the matrix representation of $SU(6)$. 
Up to an irrelevant normalization, the $U(1)$ factor is associated with the generator diag$\left(1,1,1,-1,-1,-1 \right)$.
Under the unbroken gauge group the matter content of the theory decomposes as follows:
\begin{align}
\overline{\square} &\rightarrow \overline{\square}^t_{(-1)} + \overline{\square}^b_{(1)}, \nn \\
\antim &\rightarrow \overline{\square}^t_{(2)} + \overline{\square}^b_{(-2)} + \left(\square^t, \square^b \right)_{(0)}.
\label{MatterDecompYhat}
\end{align}
%
No abelian CS term is generated for the unbroken $U(1)$. By Eq. (\ref{FITerm}), no FI term is generated either. However, non-abelian terms are generated for both $SU(3)$ factors:
\begin{align}
k^t_{eff}= \frac{1}{2}\left(4*(-1) +2*(+1) \right) = -1~, \nn \\
k^b_{eff}= \frac{1}{2}\left(4*(+1) +2*(-1) \right) = 1~.
\label{NonAbCSYhat}
\end{align}
We now argue that it is plausible that the effect of these non-abelian CS terms, combined with non-perturbative dynamics, result in the $\hat{Y}$ direction being lifted. The non-abelian CS-terms generated by the $\hat{Y}$ direction affect the equations of motions for the VEVs of the two unbroken $SU(3)$ in a manner analogous to Eqs. (\ref{DFlatCond1}-\ref{DFlatCond2}). Since turning on the directions $Y$ or $\tilde{Y}$ requires giving VEVs to some of the generators which live in these $SU(3)$ subgroups, we can see that chiral effects forbid us from turning on the $\hat{Y}$ direction when either of the other two directions are turned on. However, if we study the different regions of the CB of this theory, we can see that in all the regions where the $\hat{Y}$ operator is defined, the $\tilde{Y}$ operator is necessarily also defined. Ultimately it is the full dynamics that will determine whether along these regions of the CB the theory settles along the $\tilde{Y}$ or $\hat{Y}$ directions. The effective low energy description \eqref{WafterE} obtained through the compactification suggests that in all of these branches the theory settles to the $\tilde{Y}$ direction, and so $\hat{Y}$ gets lifted. 

It would be very interesting to gain a more detailed understanding of the underlying dynamics leading to the lifting of $\tilde{Y}$. In particular one can construct a UV free completion of the IR sector, e.g. by integrating in some extra singlets, hoping to better understand the nature of the duality under consideration here.
\footnote{ We would like to thank Ofer Aharony for emphasizing this issue to us.}
 A more detailed analysis of the different regions of the CB and their relationship to the properties of the unlifted directions will be provided in a future publication \cite{Amariti:2015sconf}.

\subsubsection*{$U(1)$ Charges}
A non-trivial check of the claimed mapping between the fields $\tilde{M}_0$, $\tilde{M}_3$ and $Y$, $A_{\mathbb{1}_{(0,-4)}}\tilde{Y}$ is the matching of the global quantum numbers on both sides of the duality. The global quantum numbers of the fields on ``the magnetic side'' are listed in Tab.~\ref{tab:3dmatter} and are inherited from those of the ``parent" 4D theory. In the 3D theory there is an extra $U(1)$ symmetry, $U(1)_4$, which originates from the diagonal generator of the flavor $SU(5)$ in 4D. On ``the electric side'', the quantum numbers of the CB parameters are inherited from the $Y_i$'s monopole operators of which they are composed. These operators acquire global symmetry charges at one-loop \cite{Aharony:1997bx} which can be computed by counting matter zero modes using the Callias index theorem \cite{Callias:1978it,deBoer:1997kr,Aharony:1997bx,Csaki:2014cwa}. We use this fact to compute the global charges of the two relevant CB parameters $Y$ and $\tilde{Y}$. As explained above, the CB splits in different regions and these operators are associated to inequivalent ones. We report the detailed calculation elsewhere \cite{Amariti:2015sconf}. 
Carefully keeping track of the splitting and the non-perturbative contributions to the super-potential, we obtain the following charge assignments:
\begin{align}
\begin{tabular}{l|ccc}
&$U(1)_{3}$&$U(1)_4$&$U(1)_{R'}$\\
\hline
$Y_1$&0&-4&10\\
$Y_{i\neq1}$&0&0&-2\\
\hline
$Y \equiv Y_1 Y_2 Y_3 Y_4 Y_5$&0&-4&2\\
$\tilde{M}_0$&0&-4&2\\
\end{tabular}
\label{TableYE}
\end{align}
The charges of $Y$ match those of the the operator $\tilde{M}_0$, as claimed.
For $\tilde{Y}$:
\begin{align}
\begin{tabular}{l|ccc}
&$U(1)_{3}$&$U(1)_4$&$U(1)_{R'}$\\
\hline
$Y_2$&6&-4&8\\
$Y_{i\neq2}$&0&0&-2\\
\hline
\rule{0pt}{3ex} $\tilde{Y} \equiv \sqrt{Y_1 Y_2^2 Y_3^2 Y_4^2 Y_5}$&6&-4&2\\
$A\tilde{Y}$&9&-4&2\\
$\tilde{M}_3$&9&-4&2\\
\end{tabular}
\label{TableYtildeE}
\end{align}
We see that while $\tilde{Y}$ by itself does not match any of the operators on ``the magnetic side'' of the theory, the dressed monopole, $A_{\mathbb{1}_{(0,-4)}}\tilde{Y}$, has the same abelian global quantum numbers as $\tilde{M}_3$. In addition, because of the $A_{\mathbb{1}_{(0,-4)}}$ field, the dressed monopole is in the $\square$ representation of the global flavor $SU(2)$, just like $\tilde{M}_3$. Since we expect the $\hat{Y}$ operator to be lifted we do not list its quantum numbers. 
%
%
%
%
%
%
%
%

\subsection{Consistency Check from the Partition Function}
Finally we discuss a further powerful check of our conjectured IR dynamics
by reducing the four dimensional superconformal index\footnote{
 The definition of the index requires only a conserved R-current; 
the theories do not necessarily have to be superconformal.}, a topological invariant quantity counting
a set of protected BPS operators in a 4D supersymmetric 
field theory, \cite{Romelsberger:2005eg,Kinney:2005ej} to the three dimensional partition function \cite{Jafferis:2010un,Hama:2010av} which is a measure of the 3D  degrees of freedom. 
Because of the technical nature of this section we will not report many of the details. They will be discussed thoroughly in \cite{Amariti:2015sconf}.


 Starting from the result of \cite{Spiridonov:2009za,Spiridonov:2014cxa} of the matching between the 4D indices in the confining case, we can show the identity of the 3D partition function for the expected 3D duals  through dimensional reduction. We observe in this process the appearance of the extra dressed monopole operators discussed above. This relation between the 4D index and 3D partition function has already been used to study the dimensional reduction of 4D dualities in \cite{Aharony:2013dha,Amariti:2014iza}. 

In our case, from the exact identity between the 4D index of the $SU(6)$ theory and the index of its confining phase,  
we obtain the relation between the  partition functions for the effective duality on $R^3 \times S^1$.
We can consider the compactified theories as \emph{effective} 3D theories with the finite size effects
from $S^1$ representing the non-perturbative dynamics. We obtain the following relation
matching the partition functions for the dual phases:
\begin{eqnarray}
\label{Zcircle}
&&
\int
\frac{  \prod_{i=1}^{6} \big(
d\sigma_i \Gamma_h(\mu_1 +\sigma_i)\prod_{\beta=1}^{5}\Gamma_h(\nu_\beta -\sigma_i) \big)
  \delta(\sum \sigma_i)}{\prod_{i<j} \Gamma_h(\pm(\sigma_i-\sigma_j))
 \prod_{\gamma=1}^{2}\Gamma_h^{-1}(\sigma_i+\sigma_j+\tau_\gamma)}
= \nonumber \\
&&
\prod_{\beta=1}^{5}\big(
 \Gamma_h(\mu_1 + \nu_\beta) \prod_{\gamma=1}^{2}\big(
 \Gamma_h(
\mu_1 + \tau_1+\tau_2+\tau_\gamma+ \nu_\beta)
 \nonumber \\
&&
\prod_{\rho=\beta+1}^{5} \!\!\!\!\Gamma_h (\tau_\gamma+ \nu_\beta+\nu_\rho) \,\,\big)\!\! 
\prod_{\rho=\beta+1}^{5} \!\!\!\!\Gamma_h(2(\tau_1+\tau_2)+ \nu_\beta+\nu_\rho) \big)\nonumber \\
&&
\prod_{i_M \in \{1,2\}}\Gamma_h( \tau_{i_1}+\tau_{i_2}+\tau_{i_3})
\end{eqnarray}
The functions $\Gamma_h$ are called hyperbolic gamma functions 
\cite{VdB}
and they represent the one loop determinants of the 
vector and matter multiplets in the computation of the partition function from localization.
The variables $\sigma_i$ are  the eigenvalues of the scalars
in the vector multiplet as in \eqref{Mario1}. 
The parameters $\mu$, $\nu$ and $\tau$ are holomorphic combinations
of the real masses for the fields and their $R$-charges. They correspond
to turning on a background gauge field for each global symmetry. 
\footnote{The real part of these parameters reproduce the weight of the representation
for each charged matter multiplet under the (non $R$) global symmetries.
The imaginary term is associated to the gauging of the $R$-symmetry and it is
proportional to the squashing parameter through the formula $\omega = i(b+1/b)$.}
Even though in the field theory analysis we fixed the R-charges as in Table \ref{tab:3dmatter}, here we consider a more general definition,
consistent with the other abelian global symmetries. 
We observe that in this case the $R$-charges can be 
treated as unconstrained. This is consistent with the absence of superpotential in the electric theory. 
This procedure allows a better identification of the zero modes carried by the (dressed) monopole operators, acting as singlets in the dual theory, in terms of the elementary fields of the electric theory.

The non-perturbative effects from the finite size of the circle generate an extra superpotential in the electric theory. It breaks an abelian symmetry which is anomalous in the 4D 
electric parent. In the dual theory this symmetry is broken by the 4D superpotential (which
is not modified by compactification).
On the partition function
this effect corresponds to a relation between the parameters  $\mu$, $\nu$ and $\tau$.
In fact the equality \eqref{Zcircle} is valid  if the parameters satisfy the relation
\begin{equation}
\mu_1 + \sum_{\beta=1}^{5} \nu_\beta + 4 \sum_{\gamma=1}^{2} \tau_\gamma = 2 \omega
\end{equation}
The equality \eqref{Zcircle} can be further reduced to the $SU(6)$ theory with four antifundamentals and 
two antisymmetrics that we studied above by a real mass flow \cite{VdB}. 
At the end of the mass flow we obtain the relation 
\begin{eqnarray}
\label{eq:Zconfining}
&&
\int
\frac{  \prod_{i=1}^{6} \big(d\sigma_i \prod_{\beta=1}^{4}
\Gamma_h(\nu_\beta -\sigma_i) \big)\delta(\sum_{i=1}^4\sigma_i)}{\prod_{i<j} \Gamma_h(\pm(\sigma_i-\sigma_j))
 \prod_{\gamma=1}^{2}\Gamma_h^{-1}(\sigma_i+\sigma_j+\tau_\gamma)}=
\nonumber \\
&&
\Gamma_h(M_{\widetilde M_0})\Gamma_h(M_{\widetilde M_3})
\prod_{i_M \in \{1,2\}}\Gamma_h( \tau_{i_1}+\tau_{i_2}+\tau_{i_3})
 \\
&&
\prod_{\beta<\rho}^{4} 
\big(
\Gamma_h(2 (\tau_1+\tau_2)+ \nu_\beta+\nu_\rho) 
\prod_{\gamma=1}^{2}\Gamma_h (\tau_\gamma + \nu_\beta+\nu_\rho) 
\big)
\nonumber
\end{eqnarray}
From which we can read off the quantum numbers of the operators in the spectrum. We find that the two terms parameterized by $M_{\widetilde M_0}$ and $M_{\widetilde M_3}$ have the same charges as the monopole $Y$ and the dressed monopole $A \tilde Y$ discussed above. This is an additional powerful check that the correct CB directions are $Y$ and $A\tilde{Y}$ as discussed previously.  More precisely we have 
$M_{\widetilde M_0} = 2\omega(1-4 \Delta_A-2\Delta_Q)-4 m_3$
and $M_{\widetilde M_3} = (\omega(2- 5 \Delta_A-4\Delta_Q)-4 m_3+9 m_4$.
The parameters $m_3$ and $m_4$ are the real masses of the abelian $U(1)_3$
and $U(1)_4$ global symmetries.
The identity (\ref{eq:Zconfining}) holds when the condition
\begin{equation}
 \sum_{\beta=1}^{4} \nu_\beta +  \sum_{\gamma=1}^{2} \tau_\gamma
=
-4 (m_3- \omega (\Delta_{\tilde Q}+ 2 \Delta_A))
\end{equation}
is imposed on the parameters. 

We conclude this section with a comment on the matching of the 3D superconformal index which could be an extra check of the relations studied in this 
section by performing the calculation on $S^2 \times S^1$ \cite{Kim:2009wb}. This can be obtained from
direct computation or from the matching between the partition on $T^2 \times S^2$ recently discussed in \cite{Benini:2015noa,Closset:2015rna}  This check can also be 
performed by following a different strategy: first one can factorize the index on the $S_b^3$ in terms of holomorphic blocks \cite{Pasquetti:2011fj},
and then glue the blocks together as explained in \cite{Beem:2012mb} to obtain the matching of the index. We leave this analysis to future investgations.

\begin{acknowledgments}

We would like to thank Ofer Aharony and Ken Intriligator for comments on the manuscript.  
A.A. is funded by the European Research Council (ERC-2012-ADG-20120216) and acknowledges
support by ANR grant 13-BS05-0001.
C.C. and N.R.L. are supported in part by the NSF grant PHY-1316222. N.R.L. is also supported in part by NSERC of Canada. MM is supported in part by DOE grant DE-SC0011784 and partially by the U.S. National Science Foundation under CAREER Grant PHY-1151392.
A.A. would like to thank The City College of New York, UC San Diego and University of Milano-Bicocca for hospitality during the final stages of
this work.

\end{acknowledgments}

\appendix 

\section{Real Mass deformation}\label{RealMass}

The real mass deformation procedure is applied on a compactified $R^3 \times S^1$ theory to decouple a flavor and get rid of the KK terms, giving us with the 3D theory with the properties described above. We go through this procedure step-by-step below.
The field content of the 4D electric and magnetic theories is given in Table~\ref{Tab:matter}.
\begin{table}
\begin{align}
\begin{tabular}{l|c|ccccc}
&$SU(6)$&$SU(2)$&$SU(5)$&$U(1)_1$&$U(1)_2$&$U(1)_R$\\
\hline
$Q$&$\square$&1&1&-5&-4&$0$\\
$\bar{Q}$&$\overline{\square}$&$1$&$\square$&1&-4&$0$\\
$A$&$\antim$&$\square$&1&0&3&$1/4$\\
\hline
$M_0 \equiv Q\bar{Q}$&&1&$\square$&-4&-8&$0$\\
$M_3 \equiv Q A^3 \bar{Q}$&&$\square$&$\square$&-4&1&3/4\\
$B_1 \equiv  A \bar{Q}^2$&&$\square$&$\antim$&2&-5&$1/4$\\
$B_3 \equiv  A^3 $&&$\symsymrep$&1&0&9&$3/4$\\
$B_4 \equiv  A^4  \bar{Q}^2$&&1&$\antim$&2&4&$1$
\end{tabular}
\nonumber 
\end{align}
\caption{Matter content of the 4D s-confining theory along with the global symmetries and the charges of the confined mesons.\label{Tab:matter}}
\end{table}

The s-confining superpotential for the mesons of the 4D theory is:
\begin{align}
W^{4D} = \dfrac{1}{\Lambda^{11}}\left(B_4^2 M_0 + B_4 M_3 B_1 + B_3 M_3 B^2_1 + B_3^2 M_0 B_1^2 \right)~.
\label{WbeforeE}
\end{align}

We now add a vector-like real mass deformation to the unique $Q$ field and the fifth flavor of $\overline{Q}$. This can be done by "fictitiously" gauging a linear combination of the diagonal generators of the flavor groups and of $U(1)_1$ such that only these flavors are charged under this combination; we can then imagine turning on a background scalar field for this gauge group, providing us with the desired real mass deformation. The mesonic fields which are left massless (and thus remain in the spectrum) under this procedure are:
\begin{itemize}
\item $B^{ab}_1$ with $a,b < 5$, which we rename $b_1$.
\item $B_3$, which we rename $b_3$.
\item $B^{ab}_4$ with $a,b < 5$ which we rename $b_4$.
\item $M^5_0 \equiv \tilde{M}_0$.
\item $M^{5}_3 \equiv \tilde{M}_3$.
\end{itemize}
These are the fields which will be part of the s-confined description of the 3D theory.

The 3D duality can now be written by applying the real mass deformation on both sides of the 4D duality. For the electric side, this consists simply of removing one flavor. For the magnetic side, the field content is reduced to those massless mesonic fields listed above and the superpotential is obtained by setting all other fields to zero in Eq. (\ref{WbeforeE}). Doing so, we obtain the matter content displayed in Table ~\ref{tab:3dmatter} and the superpotential of Eq. (\ref{WafterE}).
%

\bibliography{FDM}

\begin{thebibliography}{43}
\expandafter\ifx\csname natexlab\endcsname\relax\def\natexlab#1{#1}\fi
\expandafter\ifx\csname bibnamefont\endcsname\relax
  \def\bibnamefont#1{#1}\fi
\expandafter\ifx\csname bibfnamefont\endcsname\relax
  \def\bibfnamefont#1{#1}\fi
\expandafter\ifx\csname citenamefont\endcsname\relax
  \def\citenamefont#1{#1}\fi
\expandafter\ifx\csname url\endcsname\relax
  \def\url#1{\texttt{#1}}\fi
\expandafter\ifx\csname urlprefix\endcsname\relax\def\urlprefix{URL }\fi
\providecommand{\bibinfo}[2]{#2}
\providecommand{\eprint}[2][]{\url{#2}}

\bibitem[{\citenamefont{Aharony et~al.}(1997)\citenamefont{Aharony, Hanany,
  Intriligator, Seiberg, and Strassler}}]{Aharony:1997bx}
\bibinfo{author}{\bibfnamefont{O.}~\bibnamefont{Aharony}},
  \bibinfo{author}{\bibfnamefont{A.}~\bibnamefont{Hanany}},
  \bibinfo{author}{\bibfnamefont{K.~A.} \bibnamefont{Intriligator}},
  \bibinfo{author}{\bibfnamefont{N.}~\bibnamefont{Seiberg}}, \bibnamefont{and}
  \bibinfo{author}{\bibfnamefont{M.~J.} \bibnamefont{Strassler}},
  \bibinfo{journal}{Nucl. Phys.} \textbf{\bibinfo{volume}{B499}},
  \bibinfo{pages}{67} (\bibinfo{year}{1997}), \eprint{hep-th/9703110}.

\bibitem[{\citenamefont{de~Boer
  et~al.}(1997{\natexlab{a}})\citenamefont{de~Boer, Hori, and
  Oz}}]{deBoer:1997kr}
\bibinfo{author}{\bibfnamefont{J.}~\bibnamefont{de~Boer}},
  \bibinfo{author}{\bibfnamefont{K.}~\bibnamefont{Hori}}, \bibnamefont{and}
  \bibinfo{author}{\bibfnamefont{Y.}~\bibnamefont{Oz}},
  \bibinfo{journal}{Nucl.Phys.} \textbf{\bibinfo{volume}{B500}},
  \bibinfo{pages}{163} (\bibinfo{year}{1997}{\natexlab{a}}),
  \eprint{hep-th/9703100}.

\bibitem[{\citenamefont{Aharony}(1997)}]{Aharony:1997gp}
\bibinfo{author}{\bibfnamefont{O.}~\bibnamefont{Aharony}},
  \bibinfo{journal}{Phys.Lett.} \textbf{\bibinfo{volume}{B404}},
  \bibinfo{pages}{71} (\bibinfo{year}{1997}), \eprint{hep-th/9703215}.

\bibitem[{\citenamefont{Karch}(1997)}]{Karch:1997ux}
\bibinfo{author}{\bibfnamefont{A.}~\bibnamefont{Karch}},
  \bibinfo{journal}{Phys.Lett.} \textbf{\bibinfo{volume}{B405}},
  \bibinfo{pages}{79} (\bibinfo{year}{1997}), \eprint{hep-th/9703172}.

\bibitem[{\citenamefont{Intriligator and Seiberg}(1996)}]{Intriligator:1996mi}
\bibinfo{author}{\bibfnamefont{K.~A.} \bibnamefont{Intriligator}}
  \bibnamefont{and} \bibinfo{author}{\bibfnamefont{N.}~\bibnamefont{Seiberg}},
  \bibinfo{journal}{Phys. Lett. B} \textbf{\bibinfo{volume}{387}},
  \bibinfo{pages}{513} (\bibinfo{year}{1996}), \eprint{hep-th/9607207}.

\bibitem[{\citenamefont{de~Boer
  et~al.}(1997{\natexlab{b}})\citenamefont{de~Boer, Hori, Ooguri, and
  Oz}}]{deBoer:1996mp}
\bibinfo{author}{\bibfnamefont{J.}~\bibnamefont{de~Boer}},
  \bibinfo{author}{\bibfnamefont{K.}~\bibnamefont{Hori}},
  \bibinfo{author}{\bibfnamefont{H.}~\bibnamefont{Ooguri}}, \bibnamefont{and}
  \bibinfo{author}{\bibfnamefont{Y.}~\bibnamefont{Oz}},
  \bibinfo{journal}{Nucl.Phys.} \textbf{\bibinfo{volume}{B493}},
  \bibinfo{pages}{101} (\bibinfo{year}{1997}{\natexlab{b}}),
  \eprint{hep-th/9611063}.

\bibitem[{\citenamefont{Hanany and Witten}(1997)}]{Hanany:1996ie}
\bibinfo{author}{\bibfnamefont{A.}~\bibnamefont{Hanany}} \bibnamefont{and}
  \bibinfo{author}{\bibfnamefont{E.}~\bibnamefont{Witten}},
  \bibinfo{journal}{Nucl.Phys.} \textbf{\bibinfo{volume}{B492}},
  \bibinfo{pages}{152} (\bibinfo{year}{1997}), \eprint{hep-th/9611230}.

\bibitem[{\citenamefont{Kapustin and Strassler}(1999)}]{Kapustin:1999ha}
\bibinfo{author}{\bibfnamefont{A.}~\bibnamefont{Kapustin}} \bibnamefont{and}
  \bibinfo{author}{\bibfnamefont{M.~J.} \bibnamefont{Strassler}},
  \bibinfo{journal}{JHEP} \textbf{\bibinfo{volume}{9904}}, \bibinfo{pages}{021}
  (\bibinfo{year}{1999}), \eprint{hep-th/9902033}.

\bibitem[{\citenamefont{Dorey and Tong}(2000)}]{Dorey:2000mirror}
\bibinfo{author}{\bibfnamefont{N.}~\bibnamefont{Dorey}} \bibnamefont{and}
  \bibinfo{author}{\bibfnamefont{D.}~\bibnamefont{Tong}},
  \bibinfo{journal}{JHEP} \textbf{\bibinfo{volume}{0005}}, \bibinfo{pages}{018}
  (\bibinfo{year}{2000}), \eprint{hep-th/9911094}.

\bibitem[{\citenamefont{Borokhov
  et~al.}(2002{\natexlab{a}})\citenamefont{Borokhov, Kapustin, and
  Wu}}]{Borokhov:2002cg}
\bibinfo{author}{\bibfnamefont{V.}~\bibnamefont{Borokhov}},
  \bibinfo{author}{\bibfnamefont{A.}~\bibnamefont{Kapustin}}, \bibnamefont{and}
  \bibinfo{author}{\bibfnamefont{X.-k.} \bibnamefont{Wu}},
  \bibinfo{journal}{JHEP} \textbf{\bibinfo{volume}{0212}}, \bibinfo{pages}{044}
  (\bibinfo{year}{2002}{\natexlab{a}}), \eprint{hep-th/0207074}.

\bibitem[{\citenamefont{Borokhov
  et~al.}(2002{\natexlab{b}})\citenamefont{Borokhov, Kapustin, and
  Wu}}]{Borokhov:2002ib}
\bibinfo{author}{\bibfnamefont{V.}~\bibnamefont{Borokhov}},
  \bibinfo{author}{\bibfnamefont{A.}~\bibnamefont{Kapustin}}, \bibnamefont{and}
  \bibinfo{author}{\bibfnamefont{X.-k.} \bibnamefont{Wu}},
  \bibinfo{journal}{JHEP} \textbf{\bibinfo{volume}{0211}}, \bibinfo{pages}{049}
  (\bibinfo{year}{2002}{\natexlab{b}}), \eprint{hep-th/0206054}.

\bibitem[{\citenamefont{Intriligator and Seiberg}(2013)}]{Intriligator:2013lca}
\bibinfo{author}{\bibfnamefont{K.}~\bibnamefont{Intriligator}}
  \bibnamefont{and} \bibinfo{author}{\bibfnamefont{N.}~\bibnamefont{Seiberg}}
  (\bibinfo{year}{2013}), \eprint{1305.1633}.

\bibitem[{\citenamefont{Intriligator}(2014)}]{Intriligator:2014fda}
\bibinfo{author}{\bibfnamefont{K.}~\bibnamefont{Intriligator}},
  \bibinfo{journal}{JHEP} \textbf{\bibinfo{volume}{1410}}, \bibinfo{pages}{52}
  (\bibinfo{year}{2014}), \eprint{1406.2638}.

\bibitem[{\citenamefont{Aharony et~al.}(2013)\citenamefont{Aharony, Razamat,
  Seiberg, and Willett}}]{Aharony:2013dha}
\bibinfo{author}{\bibfnamefont{O.}~\bibnamefont{Aharony}},
  \bibinfo{author}{\bibfnamefont{S.~S.} \bibnamefont{Razamat}},
  \bibinfo{author}{\bibfnamefont{N.}~\bibnamefont{Seiberg}}, \bibnamefont{and}
  \bibinfo{author}{\bibfnamefont{B.}~\bibnamefont{Willett}}
  (\bibinfo{year}{2013}), \eprint{1305.3924}.

\bibitem[{\citenamefont{Aharony et~al.}(2015)\citenamefont{Aharony, Narayan,
  and Sharma}}]{Aharony:2015pla}
\bibinfo{author}{\bibfnamefont{O.}~\bibnamefont{Aharony}},
  \bibinfo{author}{\bibfnamefont{P.}~\bibnamefont{Narayan}}, \bibnamefont{and}
  \bibinfo{author}{\bibfnamefont{T.}~\bibnamefont{Sharma}}
  (\bibinfo{year}{2015}), \eprint{1502.00945}.

\bibitem[{\citenamefont{Pufu and Sachdev}(2013)}]{Pufu:2013eda}
\bibinfo{author}{\bibfnamefont{S.~S.} \bibnamefont{Pufu}} \bibnamefont{and}
  \bibinfo{author}{\bibfnamefont{S.}~\bibnamefont{Sachdev}},
  \bibinfo{journal}{JHEP} \textbf{\bibinfo{volume}{1309}}, \bibinfo{pages}{127}
  (\bibinfo{year}{2013}), \eprint{1303.3006}.

\bibitem[{\citenamefont{Dyer et~al.}(2013)\citenamefont{Dyer, Mezei, and
  Pufu}}]{Dyer:2013fja}
\bibinfo{author}{\bibfnamefont{E.}~\bibnamefont{Dyer}},
  \bibinfo{author}{\bibfnamefont{M.}~\bibnamefont{Mezei}}, \bibnamefont{and}
  \bibinfo{author}{\bibfnamefont{S.~S.} \bibnamefont{Pufu}}
  (\bibinfo{year}{2013}), \eprint{1309.1160}.

\bibitem[{\citenamefont{Dyer et~al.}(2015)\citenamefont{Dyer, Mezei, Pufu, and
  Sachdev}}]{Dyer:2015zha}
\bibinfo{author}{\bibfnamefont{E.}~\bibnamefont{Dyer}},
  \bibinfo{author}{\bibfnamefont{M.}~\bibnamefont{Mezei}},
  \bibinfo{author}{\bibfnamefont{S.~S.} \bibnamefont{Pufu}}, \bibnamefont{and}
  \bibinfo{author}{\bibfnamefont{S.}~\bibnamefont{Sachdev}}
  (\bibinfo{year}{2015}), \eprint{1504.00368}.

\bibitem[{\citenamefont{Csaki et~al.}(1997{\natexlab{a}})\citenamefont{Csaki,
  Schmaltz, and Skiba}}]{Csaki:1996sm}
\bibinfo{author}{\bibfnamefont{C.}~\bibnamefont{Csaki}},
  \bibinfo{author}{\bibfnamefont{M.}~\bibnamefont{Schmaltz}}, \bibnamefont{and}
  \bibinfo{author}{\bibfnamefont{W.}~\bibnamefont{Skiba}},
  \bibinfo{journal}{Phys.Rev.Lett.} \textbf{\bibinfo{volume}{78}},
  \bibinfo{pages}{799} (\bibinfo{year}{1997}{\natexlab{a}}),
  \eprint{hep-th/9610139}.

\bibitem[{\citenamefont{Csaki et~al.}(1997{\natexlab{b}})\citenamefont{Csaki,
  Schmaltz, and Skiba}}]{Csaki:1996zb}
\bibinfo{author}{\bibfnamefont{C.}~\bibnamefont{Csaki}},
  \bibinfo{author}{\bibfnamefont{M.}~\bibnamefont{Schmaltz}}, \bibnamefont{and}
  \bibinfo{author}{\bibfnamefont{W.}~\bibnamefont{Skiba}},
  \bibinfo{journal}{Phys. Rev.} \textbf{\bibinfo{volume}{D55}},
  \bibinfo{pages}{7840} (\bibinfo{year}{1997}{\natexlab{b}}),
  \eprint{hep-th/9612207}.

\bibitem[{\citenamefont{Amariti et~al.}(2015)\citenamefont{Amariti, Csaki,
  Martone, and Rey-Le~Lorier}}]{Amariti:2015sconf}
\bibinfo{author}{\bibfnamefont{A.}~\bibnamefont{Amariti}},
  \bibinfo{author}{\bibfnamefont{C.}~\bibnamefont{Csaki}},
  \bibinfo{author}{\bibfnamefont{M.}~\bibnamefont{Martone}}, \bibnamefont{and}
  \bibinfo{author}{\bibfnamefont{N.}~\bibnamefont{Rey-Le~Lorier}}
  (\bibinfo{year}{2015}), \eprint{{\it to appear}}.

\bibitem[{\citenamefont{Lee and Yi}(1997)}]{Lee:1997mo}
\bibinfo{author}{\bibfnamefont{K.}~\bibnamefont{Lee}} \bibnamefont{and}
  \bibinfo{author}{\bibfnamefont{P.}~\bibnamefont{Yi}}, \bibinfo{journal}{Phys.
  Rev.} \textbf{\bibinfo{volume}{D56}}, \bibinfo{pages}{3711}
  (\bibinfo{year}{1997}), \eprint{hep-th/9702107}.

\bibitem[{\citenamefont{Lee}(1998)}]{Lee:1998mo}
\bibinfo{author}{\bibfnamefont{K.}~\bibnamefont{Lee}}, \bibinfo{journal}{Phys.
  Lett.} \textbf{\bibinfo{volume}{B426}}, \bibinfo{pages}{323}
  (\bibinfo{year}{1998}), \eprint{hep-th/9802012}.

\bibitem[{\citenamefont{Lee and Lu}(1998)}]{Lee:1998ca}
\bibinfo{author}{\bibfnamefont{K.}~\bibnamefont{Lee}} \bibnamefont{and}
  \bibinfo{author}{\bibfnamefont{C.}~\bibnamefont{Lu}}, \bibinfo{journal}{Phys.
  Rev.} \textbf{\bibinfo{volume}{D58}}, \bibinfo{pages}{025011}
  (\bibinfo{year}{1998}), \eprint{hep-th/9802108}.

\bibitem[{\citenamefont{Kraan and van Baal}(1998{\natexlab{a}})}]{Kraan:1998pi}
\bibinfo{author}{\bibfnamefont{T.~C.} \bibnamefont{Kraan}} \bibnamefont{and}
  \bibinfo{author}{\bibfnamefont{P.}~\bibnamefont{van Baal}},
  \bibinfo{journal}{Nucl. Phys.} \textbf{\bibinfo{volume}{B533}},
  \bibinfo{pages}{627} (\bibinfo{year}{1998}{\natexlab{a}}),
  \eprint{hep-th/9805168}.

\bibitem[{\citenamefont{Kraan and van
  Baal}(1998{\natexlab{b}})}]{Kraan:1998cal}
\bibinfo{author}{\bibfnamefont{T.~C.} \bibnamefont{Kraan}} \bibnamefont{and}
  \bibinfo{author}{\bibfnamefont{P.}~\bibnamefont{van Baal}},
  \bibinfo{journal}{Phys. Lett. B} \textbf{\bibinfo{volume}{435}},
  \bibinfo{pages}{389} (\bibinfo{year}{1998}{\natexlab{b}}),
  \eprint{hep-th/9806034}.

\bibitem[{\citenamefont{Kraan and van Baal}(1999)}]{Kraan:1998mon}
\bibinfo{author}{\bibfnamefont{T.~C.} \bibnamefont{Kraan}} \bibnamefont{and}
  \bibinfo{author}{\bibfnamefont{P.}~\bibnamefont{van Baal}},
  \bibinfo{journal}{Nucl. Phys. Proc. Suppl.} \textbf{\bibinfo{volume}{73}},
  \bibinfo{pages}{554} (\bibinfo{year}{1999}), \eprint{hep-th/9808015}.

\bibitem[{\citenamefont{Csaki et~al.}(2014)\citenamefont{Csaki, Martone,
  Shirman, Tanedo, and Terning}}]{Csaki:2014cwa}
\bibinfo{author}{\bibfnamefont{C.}~\bibnamefont{Csaki}},
  \bibinfo{author}{\bibfnamefont{M.}~\bibnamefont{Martone}},
  \bibinfo{author}{\bibfnamefont{Y.}~\bibnamefont{Shirman}},
  \bibinfo{author}{\bibfnamefont{P.}~\bibnamefont{Tanedo}}, \bibnamefont{and}
  \bibinfo{author}{\bibfnamefont{J.}~\bibnamefont{Terning}},
  \bibinfo{journal}{JHEP} \textbf{\bibinfo{volume}{1408}}, \bibinfo{pages}{141}
  (\bibinfo{year}{2014}), \eprint{1406.6684}.

\bibitem[{\citenamefont{Affleck et~al.}(1982)\citenamefont{Affleck, Harvey, and
  Witten}}]{Affleck:1982as}
\bibinfo{author}{\bibfnamefont{I.}~\bibnamefont{Affleck}},
  \bibinfo{author}{\bibfnamefont{J.~A.} \bibnamefont{Harvey}},
  \bibnamefont{and} \bibinfo{author}{\bibfnamefont{E.}~\bibnamefont{Witten}},
  \bibinfo{journal}{Nucl.Phys.} \textbf{\bibinfo{volume}{B206}},
  \bibinfo{pages}{413} (\bibinfo{year}{1982}).

\bibitem[{\citenamefont{Callias}(1978)}]{Callias:1978it}
\bibinfo{author}{\bibfnamefont{C.~J.} \bibnamefont{Callias}},
  \bibinfo{journal}{Comm. Math. Phys.} \textbf{\bibinfo{volume}{63}},
  \bibinfo{pages}{213} (\bibinfo{year}{1978}).

\bibitem[{\citenamefont{Romelsberger}(2006)}]{Romelsberger:2005eg}
\bibinfo{author}{\bibfnamefont{C.}~\bibnamefont{Romelsberger}},
  \bibinfo{journal}{Nucl.Phys.} \textbf{\bibinfo{volume}{B747}},
  \bibinfo{pages}{329} (\bibinfo{year}{2006}), \eprint{hep-th/0510060}.

\bibitem[{\citenamefont{Kinney et~al.}(2007)\citenamefont{Kinney, Maldacena,
  Minwalla, and Raju}}]{Kinney:2005ej}
\bibinfo{author}{\bibfnamefont{J.}~\bibnamefont{Kinney}},
  \bibinfo{author}{\bibfnamefont{J.~M.} \bibnamefont{Maldacena}},
  \bibinfo{author}{\bibfnamefont{S.}~\bibnamefont{Minwalla}}, \bibnamefont{and}
  \bibinfo{author}{\bibfnamefont{S.}~\bibnamefont{Raju}},
  \bibinfo{journal}{Commun.Math.Phys.} \textbf{\bibinfo{volume}{275}},
  \bibinfo{pages}{209} (\bibinfo{year}{2007}), \eprint{hep-th/0510251}.

\bibitem[{\citenamefont{Jafferis}(2012)}]{Jafferis:2010un}
\bibinfo{author}{\bibfnamefont{D.~L.} \bibnamefont{Jafferis}},
  \bibinfo{journal}{JHEP} \textbf{\bibinfo{volume}{1205}}, \bibinfo{pages}{159}
  (\bibinfo{year}{2012}), \eprint{1012.3210}.

\bibitem[{\citenamefont{Hama et~al.}(2011)\citenamefont{Hama, Hosomichi, and
  Lee}}]{Hama:2010av}
\bibinfo{author}{\bibfnamefont{N.}~\bibnamefont{Hama}},
  \bibinfo{author}{\bibfnamefont{K.}~\bibnamefont{Hosomichi}},
  \bibnamefont{and} \bibinfo{author}{\bibfnamefont{S.}~\bibnamefont{Lee}},
  \bibinfo{journal}{JHEP} \textbf{\bibinfo{volume}{1103}}, \bibinfo{pages}{127}
  (\bibinfo{year}{2011}), \eprint{1012.3512}.

\bibitem[{\citenamefont{Spiridonov and Vartanov}(2011)}]{Spiridonov:2009za}
\bibinfo{author}{\bibfnamefont{V.}~\bibnamefont{Spiridonov}} \bibnamefont{and}
  \bibinfo{author}{\bibfnamefont{G.}~\bibnamefont{Vartanov}},
  \bibinfo{journal}{Commun.Math.Phys.} \textbf{\bibinfo{volume}{304}},
  \bibinfo{pages}{797} (\bibinfo{year}{2011}), \eprint{0910.5944}.

\bibitem[{\citenamefont{Spiridonov and Vartanov}(2014)}]{Spiridonov:2014cxa}
\bibinfo{author}{\bibfnamefont{V.}~\bibnamefont{Spiridonov}} \bibnamefont{and}
  \bibinfo{author}{\bibfnamefont{G.}~\bibnamefont{Vartanov}},
  \bibinfo{journal}{JHEP} \textbf{\bibinfo{volume}{1406}}, \bibinfo{pages}{062}
  (\bibinfo{year}{2014}), \eprint{1402.2312}.

\bibitem[{\citenamefont{Amariti and Klare}(2014)}]{Amariti:2014iza}
\bibinfo{author}{\bibfnamefont{A.}~\bibnamefont{Amariti}} \bibnamefont{and}
  \bibinfo{author}{\bibfnamefont{C.}~\bibnamefont{Klare}}
  (\bibinfo{year}{2014}), \eprint{1409.8623}.

\bibitem[{\citenamefont{van~de Bult}(2008)}]{VdB}
\bibinfo{author}{\bibfnamefont{F.}~\bibnamefont{van~de Bult}},
  \bibinfo{journal}{Thesis}  (\bibinfo{year}{2008}).

\bibitem[{\citenamefont{Kim}(2009)}]{Kim:2009wb}
\bibinfo{author}{\bibfnamefont{S.}~\bibnamefont{Kim}},
  \bibinfo{journal}{Nucl.Phys.} \textbf{\bibinfo{volume}{B821}},
  \bibinfo{pages}{241} (\bibinfo{year}{2009}), \eprint{0903.4172}.

\bibitem[{\citenamefont{Benini and Zaffaroni}(2015)}]{Benini:2015noa}
\bibinfo{author}{\bibfnamefont{F.}~\bibnamefont{Benini}} \bibnamefont{and}
  \bibinfo{author}{\bibfnamefont{A.}~\bibnamefont{Zaffaroni}}
  (\bibinfo{year}{2015}), \eprint{1504.03698}.

\bibitem[{\citenamefont{Closset et~al.}(2015)\citenamefont{Closset, Cremonesi,
  and Park}}]{Closset:2015rna}
\bibinfo{author}{\bibfnamefont{C.}~\bibnamefont{Closset}},
  \bibinfo{author}{\bibfnamefont{S.}~\bibnamefont{Cremonesi}},
  \bibnamefont{and} \bibinfo{author}{\bibfnamefont{D.~S.} \bibnamefont{Park}}
  (\bibinfo{year}{2015}), \eprint{1504.06308}.

\bibitem[{\citenamefont{Pasquetti}(2012)}]{Pasquetti:2011fj}
\bibinfo{author}{\bibfnamefont{S.}~\bibnamefont{Pasquetti}},
  \bibinfo{journal}{JHEP} \textbf{\bibinfo{volume}{1204}}, \bibinfo{pages}{120}
  (\bibinfo{year}{2012}), \eprint{1111.6905}.

\bibitem[{\citenamefont{Beem et~al.}(2014)\citenamefont{Beem, Dimofte, and
  Pasquetti}}]{Beem:2012mb}
\bibinfo{author}{\bibfnamefont{C.}~\bibnamefont{Beem}},
  \bibinfo{author}{\bibfnamefont{T.}~\bibnamefont{Dimofte}}, \bibnamefont{and}
  \bibinfo{author}{\bibfnamefont{S.}~\bibnamefont{Pasquetti}},
  \bibinfo{journal}{JHEP} \textbf{\bibinfo{volume}{1412}}, \bibinfo{pages}{177}
  (\bibinfo{year}{2014}), \eprint{1211.1986}.

\end{thebibliography}

\end{document}